\documentclass[acmtog,screen,nonacm]{acmart}
\setcopyright{none}
\copyrightyear{2025}
\acmYear{2025}
\acmDOI{}
\usepackage{algorithmic}
\usepackage{graphicx}
\usepackage{textcomp}
\usepackage{xcolor}
\usepackage{tabularray}

\usepackage{booktabs}
\usepackage{hyperref}
\usepackage{todonotes}
\setuptodonotes{backgroundcolor=yellow!30,linecolor=black!30}

\usepackage{tikz}
\usetikzlibrary{automata,positioning,arrows.meta,backgrounds,fit,calc}
\tikzset{arr/.style={-Latex}}
\usepackage{pgfmath}
\usepackage{tikz-cd}

\usepackage{bm} % for \circlearrowright
\usepackage[a]{esvect} % nice vector notation

\usepackage{orcidlink}

\theoremstyle{plain}
 \newtheorem{theorem}{Theorem}[section]

 \theoremstyle{definition}
 \newtheorem{definition}[theorem]{Definition}
 \newtheorem{example}[theorem]{Example}

\newcommand{\cS}{\mathcal{S}}
\newcommand{\cT}{\mathcal{T}}
\DeclareMathOperator{\Ob}{Ob}
\DeclareMathOperator{\dom}{dom}
\DeclareMathOperator{\cod}{cod}

\newcommand{\ENDO}[1]{#1^{\bm{\circlearrowright}}}

\begin{document}

\title{Bringing Algebraic Hierarchical Decompositions to Concatenative Functional Languages
}
%IEEE
% \thanks{This project was funded in part by the Kakenhi grant
%22K00015 by the Japan Society for the Promotion of Science (JSPS), titled `On progressing human understanding in the shadow of superhuman
%deep learning artificial intelligence entities' (Grant-in-Aid for Scientific
%Research type C, \url{https://kaken.nii.ac.jp/grant/KAKENHI-PROJECT-22K00015/}).}}
% \author{
% \IEEEauthorblockN{Attila Egri-Nagy}
% \IEEEauthorblockA{\textit{Human \& AI Center}\\\textit{Akita International University}\\
% Akita, Japan \\
% \orcidlinkf{0000-0001-7861-7076}}}
% \maketitle
%%%%%%%%%%%%%%

%ACM
\author{Attila Egri-Nagy}
\affiliation{\institution{Akita International University}\city{Akita-city}\country{Japan}}\email{egri-nagy@aiu.ac.jp}\orcid{0000-0001-7861-7076}
\settopmatter{printacmref=false}
\settopmatter{printfolios=true}
\renewcommand\footnotetextcopyrightpermission[1]{}
\makeatletter
\let\@authorsaddresses\@empty
\makeatother
%%%%%%%%%%%

\begin{abstract}
Programming languages tend to evolve over time to use more and more concepts from theoretical computer science.
Still, there is  a gap between programming and pure mathematics.
Not all theoretical results have realized their promising applications.
The algebraic decomposition of finite state automata (Krohn-Rhodes Theory) constructs an emulating hierarchical structure from  simpler components for any computing device.
 These decompositions provide ways to understand and control computational processes, but so far the applications were limited to theoretical investigations.
 Here, we study how to \emph{apply algebraic decompositions to programming languages}.
 We use recent results on generalizing the algebraic theory to the categorical level (from semigroups to semigroupoids) and work with the special class of concatenative functional programming languages.
As a first application of semigroupoid decompositions, we start to design a family of programming languages with an explicit semigroupoid representation.
\end{abstract}

\maketitle

%IEEE
%\begin{IEEEkeywords}
%denotational semantics, programming languages, algebraic automata theory, category theory, semigroups, semigroupoids, Krohn-Rhodes Theory, functional programming, concatenative languages
%\end{IEEEkeywords}
%%%%%%%

%\tableofcontents

\section{Introduction}

Bringing mathematics and programming closer is not a new idea.
The development of  \textsc{Fortran}, one of the earliest successful languages, started with the question ``\ldots can a machine translate a sufficiently rich mathematical language into a sufficiently economical
program \ldots ?''\cite{backus1978history}.
However, the emphasis was on expressing scientific computation, not on the mathematical properties of the language itself.
Two decades later \cite{backus78}, it was further suggested that programming should stay close to the mathematical notion of a function, opening a line of research leading to today's functional languages.
\emph{Category theory}, the study of abstract functions \cite{awodey2010category}, naturally got connected to computer science \cite{barr1995category, pierce1991basic}.
In the Categorical Abstract Machine \cite{CAM1987} this connection is made explicit.

Not all mathematics relevant to computing found a way to applications.
One prominent example is Krohn-Rhodes Theory \cite{primedecomp65}, which is about series-parallel decompositions of automata.
The theory brings together several algebraic topics: \emph{What are the basic building blocks of algebraic structures?} \emph{When are two structures similar (homomorphic), or the same (isomorphic)?}
\emph{What computation can be expressed in a structure, and what cannot?}
There are promising applications across many fields \cite{wildbook}, but the strictly automata theoretic formulations are not flexible enough for those.
One particular decomposition algorithm have been translated to categories\cite{KRTforCategories}, but it lacked software implementation.
However, recent advances opened up the space of feasible applications \cite{egrinagy2024relation,sgpoiddec,egrinagy2025minikanren}.
Leveraging this new opportunity, we directly apply semigroupoid decompositions to programming languages.

\emph{What are the potential benefits of semigroupoid decompositions of programmign languages?}
They can appear on two levels:
\begin{enumerate}
\item language design, and
\item problem solving.  
\end{enumerate}
First, a semigroupoid representation of a programming language is a mathematical object, so it gives a precise view of the structure of the language and allows comparisons between languages.
It gives a finer measure than the binary decision of computational universality (Turing-complete or not).

Second, we would like to distinguish between, and classify different solutions for a fixed problem in the same programming language.
Studying multiple solutions can be used for establishing correctness, improving efficiency, communication and teaching purposes or simply for aesthetic reasons.
We conjecture that \emph{different algorithmic solutions can be characterized by the different decompositions of the programming language.}
For instance, we would normally distinguish between recursion on a sequential collection, or folding (reducing) the same collection, since the realized computational processes are different.
On the other hand, we can argue that they are equivalent, as folding can be implemented by recursion.
We would like to clarify and decide questions like this on the algebraic level.

\subsection*{Structure of the paper}
Section \ref{sect:autdecomp} introduces the algebraic hierarchical decompositions of automata, and draws a parallel between  the motivating example of the Rubik's Cube analysis and the envisioned language decompositions.
Section \ref{sect:semigroupoids} precisely defines semigroupoids and sketches the recently developed decomposition algorithm, which enables the proposed application of programming language decompositions.
Section \ref{sect:proglang} is a brief survey of languages with the required programming language features.
The search homes in on concatenative functional languages.
Section \ref{sect:sgpoidlang} contains the preliminary work for programming language decompositions.
We close with brief conclusions and outline the next stage of this project.

\section{Hierarchical Automata  Decompositions}
\label{sect:autdecomp}

\subsection{Decomposing Automata}

 \emph{Finite state automata} (FSA) are fundamental tools in computer science \cite{HopcroftUllman79}, used in digital circuit design and compiler construction.
Without added structure for language recognition, they are essentially \emph{discrete dynamical systems}.
Thus, they can model biological systems and intelligent agents.
Algebraically, we define an automaton as a \emph{transformation semigroup} $(X,S)$, where
$X$ is a finite set of states and $S$ is a semigroup of total functions $X\rightarrow X$.
This form allows us to use tools from abstract algebra, most notably \emph{hierarchical decompositions}, which identify basic building blocks and combine them in a network where control information flows in one direction only.

\begin{figure}[h]
    \includegraphics[width=.49\textwidth]{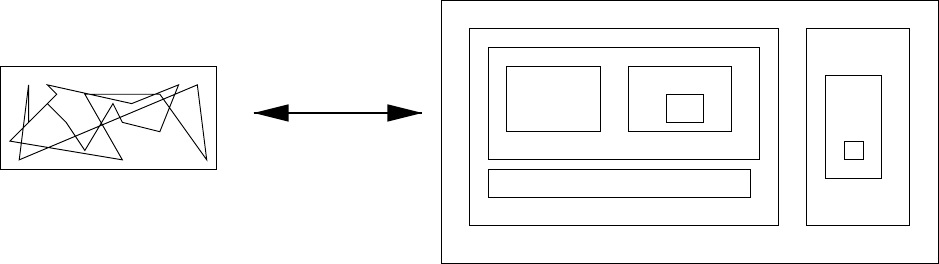}
    \caption{Schematic explanation of hierarchical decompositions. On the left we have the original system with a complicated structure. For understanding its dynamics, we build another system (on the right) with a hierarchical structure. The decomposition is potentially bigger than the original system. The hierarchical structure \emph{emulates} the original: we can ask how to do an original action in the hierarchical form. The arrow pointing from the decomposition is \emph{interpretation}: we can ask what is the meaning (in the original system) of a hierarchical action.}
    \label{fig:homomorph}
  \end{figure}
%\noindent
It is a common technique for understanding something by representing it as another object, a more familiar one, or one with a better structure (see Figure \ref{fig:homomorph}).
% In linguistics, we have cognitive metaphors for this purpose;
In mathematics, we call these homomorphisms, structure-preserving maps.
For FSA, we have \emph{emulation}: an emulating automaton can carry out at least as much computation as the emulated
one.
\emph{Decomposition} is a special form of emulation: we build an automaton from simpler ones with hierarchical (one-way) connections between the components.
The seminal result, the Krohn-Rhodes Theorem \cite{primedecomp65} states that we can always create such decompositions.
These are easy to manipulate \emph{cognitive tools} for understanding the structure and behaviour of automata, promising a wide range of applications \cite{wildbook}.

\subsection{Understanding the Rubik's Cube}

\begin{figure*}
\begin{center}
\begin{tabular}{*3l}
  \toprule
  application domain & mathematical structure & meaning of decompositions \\
  \midrule
  Rubik's Cube & symmetry group & strategies for solutions \\
  program space & semigroupoid & problem solving techniques, coding styles \\
  \bottomrule
\end{tabular}
\end{center}
\caption{The parallel between decompositions of the symmetry group of the Rubik's Cube and the decompositions of the program space of a programming language.}
\label{fig:analog}
\end{figure*}

The motivation for studying programming languages through algebraic  decompositions methodology partially came from a previous successful application of algebraic decompositions.
The symmetry group of the Rubik's Cube contains all the possibilities, the total space of arrangements a player can explore.
It is natural to use computer algebra to study permutation puzzles \cite{joyner2008adventures}, and the hierarchical decompositions are particularly good for  expressing solution strategies  \cite{2015rubik}.
A decomposition of the cube group corresponds to one particular strategy for solving.
The standard beginner method starts with making the cross on the white face of the cube.
We locate the white edge faces and move them to their positions.
This is a smaller puzzle inside the Rubik's Cube.
We only need to think about the four white edge pieces, and ignore the others.
Once the cross is done, we do not move them any more, so we are left with a smaller group, the stabilizer group of the cross.
We still move temporarily the edges of the cross, but every operation sequence will move them back.
This way we decomposed the problem into two.
Then, we can iterate this process.
Other methods choose different first goal and have accordingly different stabilizer groups.
As a more concrete example from \cite{2015rubik}, the smaller $2\times 2$ cube (Pocket Cube) can be solved by an overly systematic algorithm represented by the wreath product decomposition:
$$S_8\wr C_3 \wr S_7\wr C_3 \wr S_6\wr C_3 \wr S_5\wr C_3 \wr S_4\wr C_3 \wr S_3\wr C_3 \wr C_2\wr C_3,$$
meaning that first we solve the position of a corner regardless of other parts.
The symmetric group $S_8$ at the top level encodes this.
Then we fix the orientation of that corner by working in $C_3$, the cyclic group of order 3, and iterate the process for the remaining corners.
Another strategy is to move all the corners to the correct positions, and deal with their orientations at the same time:
$$S_8 \wr \left(\prod_{i=1}^{7}C_3\right).$$
Given a particular arrangement, the different decompositions will give different sequences of moves to solve the cube.
We are interested in these different ways of understanding the puzzle, not necessarily in the quickest solution.

\subsection{Understanding Programming Languages and Programs}

\begin{figure}
   \begin{tikzpicture}[align=center,node distance=5cm]
\node [draw](in) {I\\N\\ P\\U\\T};
\node [draw,right of=in] (out) {O\\U\\T\\P\\U\\T};
\draw (in.52) to[out=47,in=115] (1.,1.2) to[out=-180+115] (out.110);
\draw (in.52) to[out=45,in=115] (1.,1.1) to[out=-180+115] (out.115);
\draw (in.47) to[out=45,in=115] (1,1) to[out=-180+115] (out.120);
\draw (in.42) to[out=42,in=105] (1,.9) to[out=-180+105] (out.130);
\draw (in.30) to[out=35,in=143] (2,0) to[out=-180+143] (out.150);
\draw (in.25) to[out=35,in=143] (1.9,-.1) to[out=-180+143] (out.160);
\draw (in.20) to[out=35,in=143] (1.8,-.2) to[out=-180+143] (out.170);
\draw (in.-65) to[out=35,in=180] (1.5,-1) to[out=-180+180] (out.250);
\draw (in.-65) to[out=35,in=180] (1.5,-1) to[out=-180+180] (out.253);
\draw (in.-65) to[out=35,in=180] (1.5,-1) to[out=-180+180] (out.256);
\end{tikzpicture}
\caption{The algorithmic solution space is the set of computer programs solving a computational problem, i.e.~realizing a function described by input-output mappings. This schematic diagram emphasizes the space-like nature. Some solutions are closer to each other, others are more different. As programs are complex structures, we cannot measure distance with a single number, but by algebraic structures.}
\label{fig:solutionspace}
\end{figure}
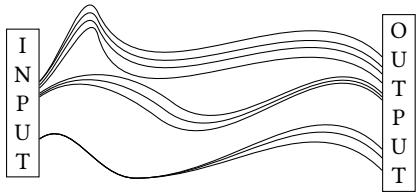

Similar to the arrangement space of the Rubik's Cube, the \emph{program space} of a language  is the set of all (syntactically valid) programs written in it.
Of course, this is a vast space, and unlike the puzzle, it is infinite.
It is more useful to think about the \emph{solution space} of a particular computational problem (see also Figure \ref{fig:solutionspace}).
For a given input-output pair we would like to know what is the set of all distinct solutions.
The basic task of sorting a sequence is an immediate example, as we have several ways of doing it with different performance characteristics.
Then, as a second layer of differences, we have stylistic variations (beyond naming). 
Sometimes individual programmers can be recognized by the style of code they write, based on the preferences for certain built-in functions and the selections of coding patterns.
  The differences can be on the level of implementation or of  the algorithm specification. Informally we can say that computations can differ by their
 intermediate results,
applied  operations,
modular structure,
or by any combination of these, shaped by language primitives, performance constraints or the programmers' experience level.

\emph{How can we detect and characterize these differences in a mathematically precise way?}
Traditionally, these questions are discussed in \emph{formal semantics} (e.g., \cite{gunter1992semantics}).
Our working hypotheses is that the algebraic decompositions of the underlying programming language gives that tool.
As the Rubik's Cube examples suggests, the solution space spanned by problem solving strategies, different coding styles can be characterized by the space of algebraic decompositions (Figure \ref{fig:analog}).

\section{Semigroupoids}
\label{sect:semigroupoids}

To benefit from the algebraic decompositions, we need to have an algebraic representation.
The FSA representation is too low-level for representing programs directly.
In particular, programming languages have types to determine what computation is admissible in given contexts.
Therefore, we need to introduce typing to automata, i.e., we generalize them to abstraction level of category theory.
Fortunately, this generalization has already been done in \cite{sgpoiddec} for simplifying decomposition algorithms.

\subsection{Definition}

Informally, we can define semigroupoids by saying that they are categories, but with the condition for identity arrows for each object removed.
Alternatively, we can start with semigroups, and restrict composition, giving rise to types.
To fix the notation, we give a complete definition.
\begin{definition}
  \label{def:sgpoid}
  A \emph{semigroupoid} consists of \emph{objects}, \emph{arrows} between the objects, and an \emph{associative} \emph{composition} operation on the arrows.
For a semigroupoid $\cS$ we use the following notation.

  \noindent\textbf{objects} The set of objects is denoted by $\Ob(\cS)$. For objects we use letters $\sigma,\tau,\rho$. We also call these (formal) \emph{types}.

  \noindent\textbf{arrows} The total set of arrows is seldom mentioned. Instead, for an object pair $(\sigma,\tau)$ we talk about the set of arrows between those objects, denoted by $\cS(\sigma,\tau)$. It can be empty, or contain one or more arrows. An arrow $a\in\cS(\sigma,\tau)$ has \emph{domain} $\dom(a)=\sigma$, and \emph{codomain} $\cod(a)=\tau$ (also called \emph{source} and \emph{target}). Alternatively, we can write $a:(\sigma,\tau)$, which reads as `$a$ has type  $(\sigma,\tau)$'.

  \noindent\textbf{composition} Two arrows $a$ and $b$ are \emph{composable} if $\cod(a)=\dom(b)$.
    The composition is denoted by concatenation $(a,b) \mapsto ab$,
    which type-checks as  $a:(\sigma,\tau)$, $b:(\tau,\rho)$, and $ab:(\sigma,\rho)$.
    Here is a diagram for composition.
\begin{center}\begin{tikzcd}
  \sigma \arrow[r,"a",arr] \arrow[rr,"ab"',bend right=15,arr] & \tau \arrow[r,"b",arr] & \rho
\end{tikzcd}
\end{center}
Note that we compose on the right as in automata theory, not on the left as in category theory, where we would say $b\circ a$ (reading as $b$ after $a$).

 \noindent\textbf{associativity} Composition should satisfy the \emph{associativity} condition: $a(bc)=(ab)c$. Consequently, the composite $abc$ is a well-defined arrow of type $(\dom(a),\cod(c))$.
\end{definition}

\begin{figure*}[ht]
\begin{center}
\begin{tblr}{Q[c,m]Q[c,m]Q[c,m]} % why not centered then?
\begin{tikzpicture}[shorten >=1pt, node distance=2cm, on grid, auto,inner sep=2pt]
    \node[draw, circle] (X)   {$\sigma$};
    \node[draw, circle] (Y) [right of=X] {$\tau$};
    \path[->,every loop/.append style=-{Latex}]
    (X) edge [arr, loop above] node {$a$} (X)
    (X) edge [arr, loop below] node {$b$} (X)
    (X) edge [arr] node {$d$} (Y)
    (X) edge [arr,bend left=42, above] node {$c$} (Y)
    (X) edge [arr,bend right=42, below] node {$e$} (Y)
    (Y) edge [arr,loop right] node {$f$} (Y);
\end{tikzpicture}
&
\begin{tblr}{
  hline{2-8}={2-8}{0.4pt,solid},
  vline{2-8}={2-8}{0.4pt,solid}}
    & $a$ & $b$ & $c$ & $d$ & $e$ & $f$ \\
$a$ & $a$ & $b$ & $c$ & $d$ & $e$ & \\
$b$ & $b$ & $a$ & $c$ & $e$ & $d$ & \\
$c$ &  &  &  &  &  &  $c$\\
$d$ &  &  &  &  &  &  $c$\\
$e$ &  &  &  &  &  &  $c$\\
$f$ &  &  &  &  &  &  $f$ \\
\end{tblr}
&
\begin{tblr}{
hline{2-5}={2-4}{0.4pt,solid},
vline{2-5}={2-4}{0.4pt,solid}}
 % this multicolumn trick is for removing the vertical lines in the header
           & $\ENDO{\sigma}$ & $\vv{\sigma\tau}$ & $\ENDO{\tau}$  \\
$\ENDO{\sigma}$ & $\ENDO{\sigma}$ & $\vv{\sigma\tau}$ &  \\
$\vv{\sigma\tau}$ &  &  & $\vv{\sigma\tau}$ \\
$\ENDO{\tau}$ &  & & $\ENDO{\tau}$ \\
\end{tblr}
\end{tblr}
\end{center}
\caption{A semigroupoid with two objects and six arrows. Diagram of objects and arrows on the left, the corresponding composition table in the middle, and the simplified composition table only with the arrow types on the right. The sequence $abaacff$ is typed as the domains match the codomains for each pair in the sequence. However, $bdef$ is not valid $d$ is not composable with $e$.}
\label{2obj6arr}
\end{figure*}
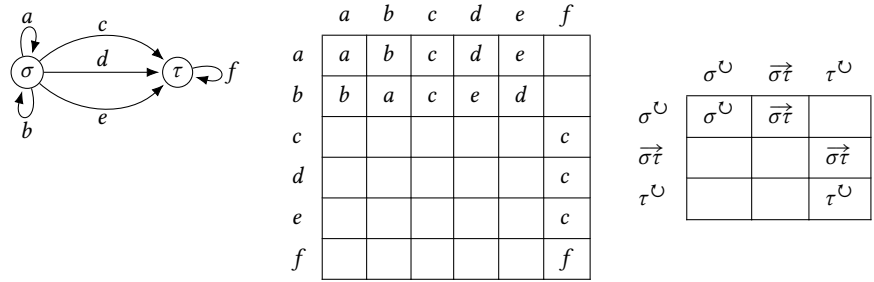

\begin{example}[Two-type semigroupoid \cite{egrinagy2025minikanren}] Fig.~\ref{2obj6arr} shows a semigroupoid with two types $\sigma,\tau$. Morphisms $a,b$ are of type $\ENDO{\sigma}$, $c,d,e$ are of type $\vv{\sigma\tau}$, and $f$ is of type $\ENDO{\tau}$.
  \label{ex:2obj-sgpoid}
\end{example}

An initial semigroupoid exploration have been carried out \cite{egrinagy2025minikanren},  finding that there can be several type structures for the same abstract semigroupoid (partially defined semigroup multiplication tables), and this underlying type structure (arrow-type semigroupoid) has central importance in studying semigroupoids.

In algebra, we often define structures by giving a set of \emph{generators}, i.e., a set of elements that yield the whole structure when compose them systematically.
For semigroupoids, the generators are arrows, forming a directed graph.
We produce the generated semigroupoid by taking the transitive closure of this graph.

\subsection{Decompositions}

The Covering Lemma method \cite{egrinagy2024relation} creates a two level decomposition of a semigroup based on a surjective morphism.
The method has been generalized to semigroupoids \cite{sgpoiddec} to make sure that the second level is also of the same kind of algebraic structure, since applying the algorithm strictly to semigroups does not necessarily give a semigroup on the second level.

The underlying idea of the decomposition algorithm is simple: we create an approximation on the top level, and recover the lost information on the bottom level.
Changes on the top level imply changes at the bottom, but not vice-versa.
Thus, the structure is hierarchical.
The algorithm to create this two level decomposition has three steps.

\noindent\textbf{collapse} A surjective morphism $\varphi: \cS\rightarrow \cT$ is the input of the decomposition algorithm. The map $\varphi$ can take several arrows to a single arrow, hence the name collapsing. The image $\cT$ is the top level of the hierarchy.

\noindent\textbf{copy} For each arrow in $a\in\cT$, we copy its preimage (everything that maps to arrow $a$). That preimage form a bottom level component, recovering all the forgotten information in $\varphi$.

\noindent\textbf{compress}  Some of the bottom level components are the same (isomorphic), so we can apply compression: store a representative once and the mappings to each copy.

Our goal is to apply this method to programming languages, thus we need to create a semigroupoid representation of a language.

\section{Decomposing Languages}
\label{sect:proglang}

In software engineering we do not work with algebraic structures directly.
We write programs in a high-level language.
Consequently, there is a gap between theory and practice.
We aim to bridge this gap by \emph{developing an experimental programming language with a semigroupoid structure}.

\begin{figure}[h]
  \begin{center}
  \begin{tabular}{l}
  infix: mathematics, most languages\\
\texttt{(2*3)*(3+4)}\\
  prefix (\textsc{Lisp}-like):\\
\texttt{(* (* 2 3) (+ 3 4))}\\ 
  postfix (stack-based):\\
  \texttt{2 3 * 3 4 + *}
  \end{tabular}
  \end{center}
  \caption{The three different ways of writing operators and their operands. Mathematics uses the infix notation. Perhaps due to its familiarity, most programming languages use the same. Languages in the \textsc{Lisp} family use prefix notation. This allows to extend the operators to $n$-ary, e.g., \texttt{(+ 1 2 3)}. It also has a uniform structure for arithmetic an function calls. Finally stack-based languages use postfix notation. This does not even need parentheses, but the arity of operators are fixed.}
  \label{fig:fixstyles}
\end{figure}

First, we have to find a suitable syntax.
Figure \ref{fig:fixstyles} shows the three  possibilities of writing operations and their operands.
The postfix, \emph{concatenative}, notation matches the sequential composition notation of semigroupoids (see Figure \ref{2obj6arr}).
It is arguable, that changing between these notations is a trivial translation.
For instance, balanced parentheses encode the behaviour of a stack.
After all, there is one abstract tree structure behind the same arithmetic expression.
However, there are other language constructs (e.g., variable declarations), that are not that trivial to translate, and we want to keep the mathematical model naturally close to the semigroupoid structure.

\subsection{Relevant Languages}

\subsubsection{FORTH}
The language \textsc{FORTH} 
\cite{moore1974forth}
was the first concatenative language (though the term `concatenative' did not exist that time).
The purpose was to have a language almost as efficient as assembly, but using English words defined by the user.
The words denote either machine code programs, or sequences of other words.
The compilation process is simple, it follows the concatenation of words.
Here is a classic conceptual example from \cite{brodie1987starting}.
\begin{verbatim}
: WASHER  WASH SPIN RINSE SPIN ;
\end{verbatim}
The new word \texttt{WASHER} is defined as the sequence of four other words -- the symbols \texttt{:} and \texttt{;} are also words.

To ensure communication between the machine code fragments (the words) a stack is used instead of formal parameter lists of functions.
To call a function, one has to put input parameters on the stack.
A program entirely consists of subroutine calls, which is called \emph{threaded code}. This allows very short program representations, since the code is just a sequence of word identifiers referring to machine code fragments.
The execution speed is faster than interpreted languages \cite{ThinkingForth}, but not as fast as register allocation optimized native code.

\subsubsection{\textsc{Lisp}}
Although it is a language with prefix operators,
we need to mention \textsc{Lisp} for two reasons.
First, the idea of \emph{quoting}, i.e., treating programs as data \cite{1960Lisp}.
Second, its 'axiomatic' nature:
building the language from a few primitives\cite{2002LispRoots}.

Quoting allows us to treat a piece of code, represented as a list, as a data structure.
Here are two expressions from a modern \textsc{Lisp}, \textsc{Clojure} \cite{Clojure2020}.
\begin{verbatim}
(count (map inc [1 2 3 4 5]))
(count '(map inc [1 2 3 4 5])
\end{verbatim}
The first line evaluates to 5, after mapping the increment function over the collection.
The second results in 3, as the code for incrementing numbers has three parts.
Code is data, thus it can be modified by the program itself.
Originally, this was thought to be an important ingredient for artificial intelligence.

Here, we are more interested in the idea of defining a languages by a few operations.
These are the `generators' of the language.
For example,
\begin{verbatim}
last == (comp first reverse)
\end{verbatim}
therefore we do not need to define an operator for getting the last element of a list, if we can reverse the list and take the first element.

\subsubsection{\textsc{Joy}}

The language \textsc{Joy}\cite{von2001joy} uses the same stack based operation as \textsc{FORTH}, but it has a radically different design philosophy, being purely functional.
It does indeed come from philosophy \cite{APLinterview}.
Unfortunately, most of the material written about \textsc{Joy} remained unpublished (available only by individuals mirroring the original homepage), and the implementations are not usable any more (due to old \textsc{C} standard).
However, the key ideas are amply described in the draft papers.

The concatenation of two programs denotes the composition of the
functions defined by the two programs.
For example, the following program
\begin{verbatim}
[1 2 3 4]  [dup *]  map
\end{verbatim}
produces the list \texttt{[1 4 9 16]} by pushing a list to the stack, then another one containing code and then applying the higher order function \texttt{map}.
Essentially, putting code into a list quotes that code. 
Programs are built of three different types of components.

\begin{itemize}
\item \textbf{literals} values that push themselves to the stack, e.g., \texttt{0},\texttt{1}, \texttt{"mango"}
\item \textbf{operators} $n$-ary operators that take $n$ values from the stack and push some result, e.g., \texttt{+}, \texttt{swap}
\item \textbf{combinators} require quoted programs on the stack, e.g., \texttt{apply}
\end{itemize}

\emph{Why a stack?}
Every function has a single argument, the stack.
Alternatively, we could consider the whole memory content of a computer as the state.
Any operation would be a function of a global state to another global state.
Function composition would revert back to the `typeless' semigroup multiplication.
This shows that computation should be 'localized' to some extent.
As another viewpoint, we can say that computation is resource-aware in the sense that variables can used only once, as in linear logic \cite{baker1994linear}.
For example, the squaring function need to duplicate $x$ to be able compute $x^2$.

\subsubsection{\textsc{Factor}}
A modern and full-featured concatenative language is \textsc{Factor} \cite{pestov2010factor}.
Among other features, it adds variable declarations for efficiency purposes.
These are beyond the scope of this paper, but it is worth noting that concatenative languages can have practical applications.

\section{A minimal semigroupoid language}
\label{sect:sgpoidlang}

It is easy to implement a concatenative programming language.
Parsing is the same as tokenizing, and the whole language can be implemented by a lookup table from tokens to stack-to-stack unary functions.
We implemented a minimal language \textsc{con-cat} \cite{concat} for the constructions below.
We differ from \textsc{Joy} in that we need to consider the list symbols \texttt{[} and \texttt{]} as separate tokens.
Reading \texttt{[} pushes an \emph{open} list to the stack, which consumes (quotes) everything that comes after that until the closing \texttt{]}, which closes the list, leaving it at the top of the stack.

\subsection{Typing rules}

Objects in a semigroupoid (category) can have many interpretations.
The simplest way to see them as `pegs' to put the morphisms on.
This view is compatible with considering them as types.

In a stack-based language, the inputs of an operator are on the stack, so their numbers and data types determine the applicability of the operator.
This is often described in  a \emph{stack effect}, e.g., for $f(x,y)=z$:
\begin{verbatim}
(x y -- z)
\end{verbatim}
We will use the stack state as types, the objects of the semigroupoid.

\subsection{Generating sets -- Partial Order of Sublanguages}
We will design and implement a succession of increasingly more capable concatenative languages, and systematically study their decompositions, i.e., adding the language elements one by one, and see what they generate.

\begin{example}
  The \texttt{swap} operator has the effect of exchanging the top of the stack and the element below it.
  The code \texttt{swap swap} will leave the stack as it was before, so it is the same as identity \texttt{id} operator.
  Thus \texttt{swap} generates the semigroupoid with the composition table below.
  \begin{center}
\begin{tabular}{c|c| c}
  & \texttt{id} & \texttt{swap} \\
  \hline
  \texttt{id}  & \texttt{id}  & \texttt{swap}\\
\hline
  \texttt{swap} & \texttt{swap} & \texttt{id}
\end{tabular}
\end{center}

This is $\mathbb{Z}_2$, the cyclic group of order 2, essentially a modulo-2 counter.
\end{example}

\subsection{Dealing with infinity}
The Krohn-Rhodes Theory is primarily a result for \emph{finite} semigroups.
In contrast, the computation space of a programming language is infinite.
There are two strategies to deal with this discrepancy.
We can have a set of atomic programs that generate only a finite program space.
Or generalize the decomposition algorithms to the infinite case.
For a specific decomposition algorithm this is possible \cite{elston_nehaniv}.
Also, the Covering Lemma method does not require finiteness.
However, there is no software implementation for the infinite case.
Therefore, the decomposition needs to be done `manually'.
It is expected that the systematic analysis of finite sublanguages will give enough understanding for carrying out a proof for the infinite case.

There are several places where infinity comes into a programming language.
\begin{itemize}
\item \textbf{numbers} Digital computers are discrete devices, but integers $\mathbb{Z}$, or even just the  natural numbers $\mathbb{N}$ are infinite sets. 
\item \textbf{stack} We can push some data item to the stack, and then another one. There is no inherent limit in pushing, thus the stack appears to be infinite.
\item \textbf{collections} Aggregate data types do not have size limits. We can always add an element to a list, or put a collection into an other one (nesting).
\item \textbf{quoting} In principle, we can quote any expression, even an already quoted one. When quoting is represented as a list, repeated quoting is the same as a nested data structure.
\end{itemize}

Numbers are easy to deal with.
We can apply finite arithmetic, modular calculations.
This is a natural choice for computers.
An 8-bit processor can only work with values from 0 to 255.
On the algebra side, the rings representing finite arithmetic, called quotient, or residue class rings, has been  decomposed hierarchically \cite{holonomyrings}. 
In other words, the language generated by \texttt{0,1,...,n-1,+,*} has been studied mathematically. 

The finiteness of the stack is a fact for all hardware implementations.
It can be of flexible size, e.g., by letting it grow backwards at the top of the heap, but then heap size makes it finite.
Handling the stack overflow error in the semigroupoid could be problematic.
Do we block pushing to the stack, or allow the bottom of the stack erased?
We can have a boolean flag included in the stack data structure, and thus in the type definition in the semigroupoid.
However, finite stack does not imply finite programming language.
Also, artificial restriction on the language, e.g., limiting the program size to fixed number of instructions, can easily violate associativity.
Without associativity, the whole algebraic framework is inapplicable.
Thus, we need to consider finite homomorphic images of infinite languages.
\begin{figure*}
\begin{center}
\begin{tblr}{Q[c,m]Q[c,m]Q[c,m]}
\begin{tikzcd}[ampersand replacement=\&]
  \& \&  \varepsilon \arrow[dll,"0"',arr] \arrow[drr,"1",arr] \&  \&  \\
  0 \arrow[dd,"0",arr]\arrow[ddr,"1",arr] \& \& \& \& 1\arrow[ddl,"1",arr]\arrow[dd,"0"',arr] \\
 \& \& \& \& \\
  00\arrow[uu,"+",arr,bend left=30] \& 01\arrow[uurrr,"+",arr] \& \& 11\arrow[uulll,"+",arr] \& 10\arrow[uu,"+"',arr,bend right=42] 
 \end{tikzcd}
  &\ \ \ \ \ \ \ &
 \begin{tikzcd}[ampersand replacement=\&]
   \& \varepsilon \arrow[d,"0\,1",arr] \& \\
   \& \{0,1\}\arrow[dl,"0\,1",arr]\arrow[dr,"0\,1"',arr] \& \\
 \{00,11\}\arrow[ur,"+",arr,bend left=30]  \&  \& \{01,10 \}\arrow[ul,"+"',arr,bend right=30] 
 \end{tikzcd}
\end{tblr}
\end{center}    
\caption{On the left, generator arrows for binary addition stack size 2 language. The types (objects) are the possible states of the stack. For example, 01 means that first 0, then 1 was pushed to the stack. The empty stack is represented by $\varepsilon$. The type $\varepsilon$ does not have an identity, so we really need a semigroupoid to model this language.  The generated semigroupoid is infinite. The missing arrows can be found by the transitive closure of the underlying graph. On the left, the compressed form.}
\label{fig:binadd}
\end{figure*}

\subsection{Binary Addition Language}

Here, we construct a very minimal language, only for binary addition.
The stack size is limited to 2.
There are two number literals \texttt{0} and \texttt{1}, and one numerical operator \texttt{+} for addition. Starting from an empty stack, the program \texttt{1 0 +} would push \texttt{1}, then \texttt{0} to the stack, then \texttt{+} pop those values, and push their sum \texttt{1} to the stack.
The program \texttt{0 1 0} is not valid, as we have a stack overflow.
Also, the program \texttt{+} is not valid due to a stack underflow error.
Figure \ref{fig:binadd} shows a set of generator arrows for this limited language.
It also shows how compression can be applied to the semigroupoid.

The corresponding semigroupoid is infinite.
It is easy to see that once the literal \texttt{0} is pushed to the stack, the program \texttt{0 +} will act as an identity on that stack state.
We can repeat \texttt{0 +} infinitely many times.
Similar loops of the generator arrows can be contracted to other types (except the empty stack), i.e., we can find directed paths starting from a type (a stack-state) that returns back to the same type.

How can we turn this structure into a finite semigroupoid?
For the Covering Lemma method, we need to give a surjective morphism.
We can try to make the first component to be a finite semigroupoid.
Thus, the decomposition divides the infinite into two parts, one finite, and another infinite.
We might implement the identity on the stack state of containing a single \texttt{0}, but we could say that we are not interested in the details, just the fact there is an identity operation on that type.
On the top level of the hierarchy, we record the existence of the identity arrow, and on the second level we include all the infinite ways to realize that arrow.

One possible finite image is the \emph{arrow-type semigroupoid} from \cite{egrinagy2025minikanren}.
That surjective morphism sends the set of arrows between two types to a single arrow.
In a way, this gives a primitive type-checker.
The second level of the decomposition then would comprise of all these lost details, namely all the possible programs that take the source type to the target type.

Another way to have a finite representation is to consider the transformations of the global state space.
In Figure \ref{fig:binadd}, we use the same label for several arrows, though in the semigroupoid those arrows are different, and the connection is external (made on the semantic level).
If we take the stack states in a fixed  order $\varepsilon, 0,1,00,01,11,10$ (just top to bottom, left to right enumeration in the diagram), and encode by their index,
\begin{center}
\begin{tblr}{c|c|c|c|c|c|c}
$\varepsilon$ & 0 & 1 & 00 & 01 & 11& 10\\
\hline
  0 & 1 & 2 & 3 & 4 & 5 & 6
\end{tblr}  
\end{center}
then \texttt{+} defines the mappings $3\mapsto 1$, $4\mapsto 2$, $5\mapsto 1$, and $6\mapsto 2$.
For the remaining states it is undefined, since the \texttt{+} operator requires two number literals on the stack.
We get the partial transformations \texttt{+}$[\_,\_,\_,1,2,1,2]$, \texttt{0}$[1,3,6,\_,\_,\_,\_]$, \texttt{1}$[2,4,5\_,\_,\_,\_,]$.
This gives a partial transformation representation, which can easily be translated into a transformation representation (by adding an extra sink state to represent the undefined state).
They generate a semigroup with 21 elements.
When doing a complete (holonomy) decomposition, we get $\mathbb{Z}_2$ among the components, coming from the binary addition.
It is always possible to have such a partial transformation representation, but the state space grows fast.

\section{Conclusion}

We described a way to bring the algebraic hierarchical decompositions to programming languages.
This is the first step in the long-term project of building a programming language, piece by piece, maintaining an explicit algebraic understanding.
This plan requires modifications in automata theory (generalizing the algebraic representation from semigroups to semigroupoids) and restricting to the special class of concatenative languages to match the linear syntax of automata input words.
The main ideas can be summarized by the following points.
\begin{itemize}
\item Concatenative languages have the same syntax as semigroupoid composition.  
\item Stack effects define types (objects of the semigroupoid), giving categorical semantics.
\item We can manage infinity by mapping an infinite structure down to a finite one.
\end{itemize}
These enable us to feed a programming language into the decomposition algorithms, giving a  new type of mathematical understanding of computer programs.
The next task is the define a concatenative language in a \emph{modular} way.
Since a full featured language  is too big to decompose, we need to add literals, operators and combinators one by one as generators.
The source code for this project is available at \cite{concat}.

\begin{acks}
This project was funded in part by the Kakenhi grant
22K00015 by the Japan Society for the Promotion of Science (JSPS), titled `On progressing human understanding in the shadow of superhuman
deep learning artificial intelligence entities' (Grant-in-Aid for Scientific
Research type C, \url{https://kaken.nii.ac.jp/grant/KAKENHI-PROJECT-22K00015/}).
\end{acks}

\bibliographystyle{ACM-Reference-Format}
\bibliography{../coords.bib}


\begin{thebibliography}{29}

%%% the Web address.

\ifx \showCODEN    \undefined \def \showCODEN     #1{\unskip}     \fi
\ifx \showISBNx    \undefined \def \showISBNx     #1{\unskip}     \fi
\ifx \showISBNxiii \undefined \def \showISBNxiii  #1{\unskip}     \fi
\ifx \showISSN     \undefined \def \showISSN      #1{\unskip}     \fi
\ifx \showLCCN     \undefined \def \showLCCN      #1{\unskip}     \fi
\ifx \shownote     \undefined \def \shownote      #1{#1}          \fi
\ifx \showarticletitle \undefined \def \showarticletitle #1{#1}   \fi
\ifx \showURL      \undefined \def \showURL       {\relax}        \fi
% invisible to TeX
\providecommand\bibfield[2]{#2}
\providecommand\bibinfo[2]{#2}
\providecommand\natexlab[1]{#1}
\providecommand\showeprint[2][]{arXiv:#2}

\bibitem[Apter(2004)]%
        {APLinterview}
\bibfield{author}{\bibinfo{person}{Stevan Apter}.}
  \bibinfo{year}{2004}\natexlab{}.
\newblock \showarticletitle{A Conversation with {M}anfred von {T}hun}.
\newblock \bibinfo{journal}{\emph{VECTOR, Journal of the British APL
  Association}} \bibinfo{volume}{20}, \bibinfo{number}{3}
  (\bibinfo{year}{2004}).
\newblock
\newblock
\shownote{\url{https://archive.vector.org.uk/art10000350}}.


\bibitem[Awodey(2010)]%
        {awodey2010category}
\bibfield{author}{\bibinfo{person}{S. Awodey}.}
  \bibinfo{year}{2010}\natexlab{}.
\newblock \bibinfo{booktitle}{\emph{Category Theory}}.
\newblock \bibinfo{publisher}{OUP Oxford}.
\newblock
\showISBNx{9780199587360}
\showLCCN{2010483708}


\bibitem[Backus(1978a)]%
        {backus78}
\bibfield{author}{\bibinfo{person}{John Backus}.}
  \bibinfo{year}{1978}\natexlab{a}.
\newblock \showarticletitle{Can programming be liberated from the von Neumann
  style? a functional style and its algebra of programs}.
\newblock \bibinfo{journal}{\emph{Commun. ACM}} \bibinfo{volume}{21},
  \bibinfo{number}{8} (\bibinfo{date}{Aug.} \bibinfo{year}{1978}),
  \bibinfo{pages}{613–641}.
\newblock
\showISSN{0001-0782}
\href{https://doi.org/10.1145/359576.359579}{doi:\nolinkurl{10.1145/359576.359579}}


\bibitem[Backus(1978b)]%
        {backus1978history}
\bibfield{author}{\bibinfo{person}{John Backus}.}
  \bibinfo{year}{1978}\natexlab{b}.
\newblock \showarticletitle{The history of {F}ortran {I}, {II}, and {III}}.
\newblock \bibinfo{journal}{\emph{ACM Sigplan Notices}} \bibinfo{volume}{13},
  \bibinfo{number}{8} (\bibinfo{year}{1978}), \bibinfo{pages}{165--180}.
\newblock


\bibitem[Baker(1994)]%
        {baker1994linear}
\bibfield{author}{\bibinfo{person}{Henry~G Baker}.}
  \bibinfo{year}{1994}\natexlab{}.
\newblock \showarticletitle{Linear logic and permutation stacks-the Forth shall
  be first}.
\newblock \bibinfo{journal}{\emph{ACM Sigarch Computer Architecture News}}
  \bibinfo{volume}{22}, \bibinfo{number}{1} (\bibinfo{year}{1994}),
  \bibinfo{pages}{34--43}.
\newblock


\bibitem[Barr and Wells(1995)]%
        {barr1995category}
\bibfield{author}{\bibinfo{person}{M. Barr} {and} \bibinfo{person}{C. Wells}.}
  \bibinfo{year}{1995}\natexlab{}.
\newblock \bibinfo{booktitle}{\emph{Category Theory for Computing Science}}.
\newblock Number v. 1 in \bibinfo{series}{Prentice-Hall international series in
  computer science}. \bibinfo{publisher}{Prentice Hall}.
\newblock
\showISBNx{9780133238099}
\showLCCN{95004728}


\bibitem[Brodie(1987)]%
        {brodie1987starting}
\bibfield{author}{\bibinfo{person}{Leo Brodie}.}
  \bibinfo{year}{1987}\natexlab{}.
\newblock \bibinfo{booktitle}{\emph{Starting {F}orth}}.
\newblock \bibinfo{publisher}{Prentice-Hall, Inc.}
\newblock


\bibitem[Brodie(2004)]%
        {ThinkingForth}
\bibfield{author}{\bibinfo{person}{L. Brodie}.}
  \bibinfo{year}{2004}\natexlab{}.
\newblock \bibinfo{booktitle}{\emph{Thinking Forth}}.
\newblock \bibinfo{publisher}{Punchy Publishing}.
\newblock
\showISBNx{9780976458708}
\urldef\tempurl%
\url{https://www.forth.com/forth-books/}
\showURL{%
\tempurl}


\bibitem[Cousineau et~al\mbox{.}(1987)]%
        {CAM1987}
\bibfield{author}{\bibinfo{person}{G. Cousineau}, \bibinfo{person}{P.-L.
  Curien}, {and} \bibinfo{person}{M. Mauny}.} \bibinfo{year}{1987}\natexlab{}.
\newblock \showarticletitle{The categorical abstract machine}.
\newblock \bibinfo{journal}{\emph{Science of Computer Programming}}
  \bibinfo{volume}{8}, \bibinfo{number}{2} (\bibinfo{year}{1987}),
  \bibinfo{pages}{173--202}.
\newblock
\showISSN{0167-6423}
\href{https://doi.org/10.1016/0167-6423(87)90020-7}{doi:\nolinkurl{10.1016/0167-6423(87)90020-7}}


\bibitem[Egri-Nagy(2025)]%
        {concat}
\bibfield{author}{\bibinfo{person}{Attila Egri-Nagy}.}
  \bibinfo{year}{2025}\natexlab{}.
\newblock \bibinfo{booktitle}{\emph{{\textsc{con\-cat} functional concatenative
  programming language with an explicit semigroupoid representations}}}.
\newblock
\newblock
\shownote{\href{https://codeberg.org/egri-nagy/con-cat}{{codeberg.org/egri-nagy/con-cat}}}.


\bibitem[Egri-Nagy and Nehaniv(2007)]%
        {holonomyrings}
\bibfield{author}{\bibinfo{person}{A. Egri-Nagy} {and} \bibinfo{person}{C.~L.
  Nehaniv}.} \bibinfo{year}{2007}\natexlab{}.
\newblock \showarticletitle{{Finite Residue Class Rings of Integers Modulo $n$
  from the Viewpoint of Global Semigroup Theory}}. In
  \bibinfo{booktitle}{\emph{{Proceedings of International Conference on
  Semigroups and Languages 2005 In Honour of the 65th birthday of Donald B.\
  McAlister}}}. \bibinfo{publisher}{World Scientific Press},
  \bibinfo{pages}{66--83}.
\newblock


\bibitem[Egri-Nagy and Nehaniv(2015)]%
        {2015rubik}
\bibfield{author}{\bibinfo{person}{Attila Egri-Nagy} {and}
  \bibinfo{person}{Chrystopher~L. Nehaniv}.} \bibinfo{year}{2015}\natexlab{}.
\newblock \showarticletitle{Computational Understanding and Manipulation of
  Symmetries}. In \bibinfo{booktitle}{\emph{Artificial Life and Computational
  Intelligence}}, \bibfield{editor}{\bibinfo{person}{Stephan~K. Chalup},
  \bibinfo{person}{Alan~D. Blair}, {and} \bibinfo{person}{Marcus Randall}}
  (Eds.). \bibinfo{publisher}{Springer International Publishing},
  \bibinfo{address}{Cham}, \bibinfo{pages}{17--30}.
\newblock
\showISBNx{978-3-319-14803-8}


\bibitem[Egri-Nagy and Nehaniv(2024)]%
        {egrinagy2024relation}
\bibfield{author}{\bibinfo{person}{Attila Egri-Nagy} {and}
  \bibinfo{person}{Chrystopher~L. Nehaniv}.} \bibinfo{year}{2024}\natexlab{}.
\newblock \showarticletitle{From Relation to Emulation and Interpretation:
  Computer Algebra Implementation of the Covering Lemma for Finite
  Transformation Semigroups}. In \bibinfo{booktitle}{\emph{Implementation and
  Application of Automata}},
  \bibfield{editor}{\bibinfo{person}{Szil{\'a}rd~Zsolt Fazekas}} (Ed.).
  \bibinfo{publisher}{Springer Nature Switzerland}, \bibinfo{address}{Cham},
  \bibinfo{pages}{138--152}.
\newblock
\showISBNx{978-3-031-71112-1}
\href{https://doi.org/10.1007/978-3-031-71112-1_10}{doi:\nolinkurl{10.1007/978-3-031-71112-1_10}}


\bibitem[Egri-Nagy and Nehaniv(2025a)]%
        {egrinagy2025minikanren}
\bibfield{author}{\bibinfo{person}{Attila Egri-Nagy} {and}
  \bibinfo{person}{Chrystopher~L. Nehaniv}.} \bibinfo{year}{2025}\natexlab{a}.
\newblock \bibinfo{title}{Computational Exploration of Finite Semigroupoids}.
\newblock
\showeprint[arxiv]{2509.00837}~[cs.FL]
\urldef\tempurl%
\url{https://arxiv.org/abs/2509.00837}
\showURL{%
\tempurl}
\newblock
\shownote{miniKanren 2025 Workshop at ICPFL/SPLASH'25}.


\bibitem[Egri-Nagy and Nehaniv(2025b)]%
        {sgpoiddec}
\bibfield{author}{\bibinfo{person}{Attila Egri-Nagy} {and}
  \bibinfo{person}{Chrystopher~L. Nehaniv}.} \bibinfo{year}{2025}\natexlab{b}.
\newblock \bibinfo{title}{Representation Independent Decompositions of
  Computation}.
\newblock
\showeprint[arxiv]{2504.04660}~[math.GR]
\urldef\tempurl%
\url{https://arxiv.org/abs/2504.04660}
\showURL{%
\tempurl}


\bibitem[Elston and Nehaniv(2002)]%
        {elston_nehaniv}
\bibfield{author}{\bibinfo{person}{Gillian~Z. Elston} {and}
  \bibinfo{person}{Chrystopher~L. Nehaniv}.} \bibinfo{year}{2002}\natexlab{}.
\newblock \showarticletitle{{Holonomy Embedding of Arbitrary Stable
  Semigroups}}.
\newblock \bibinfo{journal}{\emph{International Journal of Algebra and
  Computation}} \bibinfo{volume}{12}, \bibinfo{number}{6}
  (\bibinfo{year}{2002}), \bibinfo{pages}{791--810}.
\newblock


\bibitem[Graham(2002)]%
        {2002LispRoots}
\bibfield{author}{\bibinfo{person}{Paul Graham}.}
  \bibinfo{year}{2002}\natexlab{}.
\newblock \bibinfo{title}{The Roots of Lisp}.
\newblock
  \bibinfo{howpublished}{\url{https://www.paulgraham.com/rootsoflisp.html}}.
\newblock


\bibitem[Gunter(1992)]%
        {gunter1992semantics}
\bibfield{author}{\bibinfo{person}{C.A. Gunter}.}
  \bibinfo{year}{1992}\natexlab{}.
\newblock \bibinfo{booktitle}{\emph{Semantics of Programming Languages:
  Structures and Techniques}}.
\newblock \bibinfo{publisher}{MIT Press}.
\newblock
\showISBNx{9780262570954}
\showLCCN{92010172}


\bibitem[Hickey(2020)]%
        {Clojure2020}
\bibfield{author}{\bibinfo{person}{Rich Hickey}.}
  \bibinfo{year}{2020}\natexlab{}.
\newblock \showarticletitle{A History of {C}lojure}.
\newblock \bibinfo{journal}{\emph{Proc. ACM Program. Lang.}}
  \bibinfo{volume}{4}, \bibinfo{number}{HOPL}, Article \bibinfo{articleno}{71}
  (\bibinfo{date}{June} \bibinfo{year}{2020}), \bibinfo{numpages}{46}~pages.
\newblock
\href{https://doi.org/10.1145/3386321}{doi:\nolinkurl{10.1145/3386321}}


\bibitem[Hopcroft and Ullman(1979)]%
        {HopcroftUllman79}
\bibfield{author}{\bibinfo{person}{John~E. Hopcroft} {and}
  \bibinfo{person}{Jeff~D. Ullman}.} \bibinfo{year}{1979}\natexlab{}.
\newblock \bibinfo{booktitle}{\emph{Introduction to Automata Theory, Languages,
  and Computation}}.
\newblock \bibinfo{publisher}{Addison-Wesley}.
\newblock


\bibitem[Joyner(2008)]%
        {joyner2008adventures}
\bibfield{author}{\bibinfo{person}{D. Joyner}.}
  \bibinfo{year}{2008}\natexlab{}.
\newblock \bibinfo{booktitle}{\emph{Adventures in Group Theory: Rubik's Cube,
  Merlin's Machine, and Other Mathematical Toys}}.
\newblock \bibinfo{publisher}{Johns Hopkins University Press}.
\newblock
\showISBNx{9780801897269}
\showLCCN{2008011322}


\bibitem[Krohn and Rhodes(1965)]%
        {primedecomp65}
\bibfield{author}{\bibinfo{person}{Kenneth Krohn} {and} \bibinfo{person}{John
  Rhodes}.} \bibinfo{year}{1965}\natexlab{}.
\newblock \showarticletitle{{Algebraic Theory of Machines. {I}. {P}rime
  Decomposition Theorem for Finite Semigroups and Machines}}.
\newblock \bibinfo{journal}{\emph{Trans. Amer. Math. Soc.}}
  \bibinfo{volume}{116} (\bibinfo{date}{April} \bibinfo{year}{1965}),
  \bibinfo{pages}{450--464}.
\newblock


\bibitem[McCarthy(1960)]%
        {1960Lisp}
\bibfield{author}{\bibinfo{person}{John McCarthy}.}
  \bibinfo{year}{1960}\natexlab{}.
\newblock \showarticletitle{Recursive functions of symbolic expressions and
  their computation by machine, {P}art {I}}.
\newblock \bibinfo{journal}{\emph{Commun. ACM}} \bibinfo{volume}{3},
  \bibinfo{number}{4} (\bibinfo{date}{April} \bibinfo{year}{1960}),
  \bibinfo{pages}{184–195}.
\newblock
\showISSN{0001-0782}
\href{https://doi.org/10.1145/367177.367199}{doi:\nolinkurl{10.1145/367177.367199}}


\bibitem[Moore(1974)]%
        {moore1974forth}
\bibfield{author}{\bibinfo{person}{Charles~H Moore}.}
  \bibinfo{year}{1974}\natexlab{}.
\newblock \showarticletitle{FORTH: a new way to program a mini computer}.
\newblock \bibinfo{journal}{\emph{Astronomy and Astrophysics Supplement, Vol.
  15, p. 497}}  \bibinfo{volume}{15} (\bibinfo{year}{1974}),
  \bibinfo{pages}{497}.
\newblock


\bibitem[Pestov et~al\mbox{.}(2010)]%
        {pestov2010factor}
\bibfield{author}{\bibinfo{person}{Sviatoslav Pestov}, \bibinfo{person}{Daniel
  Ehrenberg}, {and} \bibinfo{person}{Joe Groff}.}
  \bibinfo{year}{2010}\natexlab{}.
\newblock \showarticletitle{Factor: A dynamic stack-based programming
  language}.
\newblock \bibinfo{journal}{\emph{Acm Sigplan Notices}} \bibinfo{volume}{45},
  \bibinfo{number}{12} (\bibinfo{year}{2010}), \bibinfo{pages}{43--58}.
\newblock


\bibitem[Pierce(1991)]%
        {pierce1991basic}
\bibfield{author}{\bibinfo{person}{B.C. Pierce}.}
  \bibinfo{year}{1991}\natexlab{}.
\newblock \bibinfo{booktitle}{\emph{Basic Category Theory for Computer
  Scientists}}.
\newblock \bibinfo{publisher}{MIT Press}.
\newblock
\showISBNx{9780262660716}
\showLCCN{91008489}


\bibitem[Rhodes(2009)]%
        {wildbook}
\bibfield{author}{\bibinfo{person}{John Rhodes}.}
  \bibinfo{year}{2009}\natexlab{}.
\newblock \bibinfo{booktitle}{\emph{{Applications of Automata Theory and
  Algebra via the Mathematical Theory of Complexity to Biology, Physics,
  Psychology, Philosophy, and Games}}}.
\newblock \bibinfo{publisher}{World Scientific Press}.
\newblock
\newblock
\shownote{Foreword by Morris W.\ Hirsch, edited by Chrystopher L.\ Nehaniv
  (Original version: UC Berkeley, Mathematics Library, 1971)}.


\bibitem[Von~Thun and Thomas(2001)]%
        {von2001joy}
\bibfield{author}{\bibinfo{person}{Manfred Von~Thun} {and}
  \bibinfo{person}{Reuben Thomas}.} \bibinfo{year}{2001}\natexlab{}.
\newblock \showarticletitle{Joy: Forth’s functional cousin}. In
  \bibinfo{booktitle}{\emph{Proceedings of the 17th EuroForth Conference}}.
\newblock


\bibitem[Wells(1980)]%
        {KRTforCategories}
\bibfield{author}{\bibinfo{person}{Charles Wells}.}
  \bibinfo{year}{1980}\natexlab{}.
\newblock \showarticletitle{A {K}rohn-{R}hodes Theorem for categories}.
\newblock \bibinfo{journal}{\emph{Journal of Algebra}}  \bibinfo{volume}{64}
  (\bibinfo{year}{1980}), \bibinfo{pages}{37--45}.
\newblock
\href{https://doi.org/10.1016/0021-8693(80)90130-1}{doi:\nolinkurl{10.1016/0021-8693(80)90130-1}}


\end{thebibliography}

\end{document}